\documentclass[prl,twocolumn,showpacs,superscriptaddress]{revtex4}
\usepackage{epsf}
\usepackage{graphicx}
\usepackage{bm}
\usepackage[T1]{fontenc}
\usepackage[latin1]{inputenc}
\usepackage{times}
\usepackage{ulem}
\usepackage{color}

\begin{document}

\title{Recurrent scattering and memory effect at the Anderson localization transition}
\author{A. Aubry}
\affiliation{Institut Langevin, ESPCI ParisTech, CNRS UMR 7587, Universit\'{e} Denis Diderot - Paris 7, 1 rue Jussieu, 75005 Paris, France}
\author{L. A. Cobus}
\affiliation{Department of Physics and Astronomy, University of Manitoba, Winnipeg, Manitoba R3T 2N2, Canada}
\author{S. E. Skipetrov}
\affiliation{Universit\'{e} Grenoble 1/CNRS, Laboratoire de Physique et Mod\'elisation des
Milieux Condens\'es UMR 5493,\\ B.P. 166, 38042 Grenoble, France}
\author{B. A. van Tiggelen}
\affiliation{Universit\'{e} Grenoble 1/CNRS, Laboratoire de Physique et Mod\'elisation des
Milieux Condens\'es UMR 5493,\\ B.P. 166, 38042 Grenoble, France}
\author{A. Derode}
\affiliation{Institut Langevin, ESPCI ParisTech, CNRS UMR 7587, Universit\'{e} Denis Diderot - Paris 7, 1 rue Jussieu, 75005 Paris, France}
\author{J. H. Page}
\affiliation{Department of Physics and Astronomy, University of Manitoba, Winnipeg, Manitoba R3T 2N2, Canada}

\date{\today}

\begin{abstract}
We report on ultrasonic measurements of the propagation operator in a strongly scattering mesoglass. The backscattered field is shown to display a deterministic spatial coherence due to a remarkably large memory effect induced by long recurrent trajectories. Investigation of the recurrent scattering contribution directly yields the probability for a wave to come back close to its starting spot. The decay of this quantity with time is shown to change dramatically near the Anderson localization transition. The singular value decomposition of the propagation operator reveals the dominance of very intense recurrent scattering paths near the mobility edge.
 \end{abstract}


\maketitle
In a disordered medium, a classical approach is to consider the trajectory followed by a wave as a Brownian random walk. After a few scattering events, the spatio-temporal evolution of the mean intensity is governed by the diffusion equation. The relevant parameters are the scattering mean free path $l_s$, the transport mean free path $l^*$ and the Boltzmann diffusion coefficient $D_B$.  However, this classical picture neglects interference effects that may resist the influence of disorder. In particular, constructive interference between reciprocal multiple scattering (MS) paths enhances the probability for a wave to come back close to its starting point as compared to classical predictions: this phenomenon is known as weak localization.
Hence, interference can slow down and eventually stop the diffusion process, giving rise to Anderson localization (AL) \cite{anderson,abrahams,Vollhardt,skipetrov,Lagendijk}: instead of spreading diffusely from the source, a wave packet remains localized in its vicinity on a length scale given by the localization length $\xi$. The transition at a mobility edge between diffuse and localized behavior is predicted to exist only in three-dimensional (3D) media and occurs when the scattering is sufficiently strong, \textit{i.e.} when $k l_s \sim 1$ (with $k$ the wave number in the scattering medium) \cite{abrahams}. Several experiments in optics have shown deviations from diffuse behavior in 3D strongly scattering samples \cite{wiersma,storzer,dogariu}. However, the most direct proof of 3D classical wave localization was first established in acoustics, by observing the transverse confinement of energy in a \textit{mesoglass} consisting of an elastic network of aluminium beads \cite{hu}. More recently, this transverse confinement method \cite{hu} has also been used in optics \cite{sperling}.

\begin{figure}[htbp]
\center
\includegraphics[width=6cm]{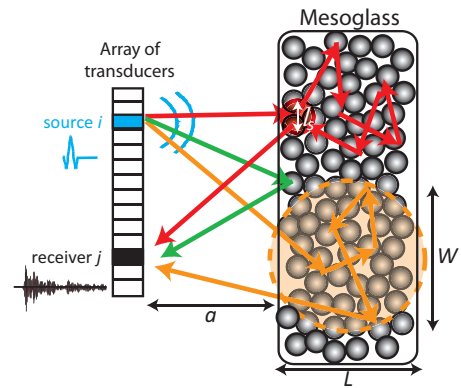}
\caption{Experimental setup.
Examples of a single scattering path (green arrows), of a RS path (red arrows) and of a conventional MS path (orange arrows) are drawn.}
\label{fig1}
\end{figure}

In this Letter, we investigate some new mesoscopic aspects of AL, taking advantage of ultrasonic technology. More precisely, we adopt a matrix formalism which is particularly appropriate since the ultrasonic wave field can be controlled by an array of independent transducers acting both as sources and receivers (Fig~.\ref{fig1}). The Green's functions $K_{ij}$, obtained by emitting a wave from an array element $i$ and recording the backscattered field at an element $j$, constitute the propagation matrix $\mathbf{K}$. This matrix provides many fundamental insights into the medium under investigation. One can, for instance, extract from $\mathbf{K}$ the single- and multiple-scattering components \cite{aubry2009random,aubry_jap,aubry_jasa}, the diffusive halo and the coherent backscattering cone \cite{aubry2}, or even determine the open scattering channels \cite{popoff2010measuring,choi}.

In this study, the propagation matrix is investigated in the strongly scattering regime ($kl_s \sim 1$). Surprisingly, the backscattered field displays a deterministic coherence along the antidiagonals of $\mathbf{K}$ in a way similar to single scattering \cite{aubry2009random}, but at much longer times of flight. We argue that recurrent scattering (RS) paths account for this long-range coherence and establish the link with the memory effect \cite{freund,feng}. By RS paths, we mean any scattering path that begins and ends at positions separated by less than one mean free path. A matrix method is then applied to extract the RS contribution from the backscattered wave field. This yields a quantitative measurement of the probability for a wave to return to its starting spot. A slowing down of the decay of this return probability with time is observed near the localization transition, in qualitative agreement with theory \cite{skipetrov}. At the mobility edge, the measured decay is actually slower than anticipated from the self-consistent (SC) theory of localization \cite{skipetrov}. This surprising behavior coincides with the emergence of very intense RS paths that dominate the singular value spectrum of $\mathbf{K}$. These observations offer new insights into key aspects of 3D AL that have not been accessible to experimental investigation previously. 

Our random scattering sample is a mesoglass similar to those used in a previous work \cite{hu}. It is made from 3.93 mm mean diameter aluminium beads brazed together at a volume fraction of approximately 55 \%. The cross-section ($230 \times 250$ mm$^2$) of the slab-shaped sample is much larger than its thickness $L = 25$ mm. Unlike the samples in Ref.~\onlinecite{hu}, there is a polydispersity of about 20 \% in the size of the beads, and different brazing conditions resulted in stronger elastic bonds between the beads, thereby modifying the scattering properties of the sample. From measurements at 0.9 MHz of the coherent pulse crossing the sample \cite{page}, we estimate the longitudinal phase velocity $v_p \approx 2.8$ mm$/\mu$s and the mean free path $l_s \approx 1.3$ mm \cite{page}. This leads to a product of wave number $k$ and mean free path $kl_s \lesssim 3$.
Above 1 MHz, the coherent pulse becomes too small to measure accurately in this thick sample, consistent with even stronger scattering and smaller values of $kl_s$.  By performing transverse confinement measurements \cite{hu}, we find that the waves are localized between 
1.2 and 1.25 MHz (mobility edges), with $\xi < L$ in the middle of this band.

The experimental setup (Fig.~\ref{fig1}) uses $N = 64$ elements of a linear ultrasonic array in the 1--2 MHz frequency range. The array pitch $p$ is 0.5 mm. The array is immersed in water, at a distance $a=182$ mm from the waterproofed sample. $N^2$ time-dependent responses are measured by sending a pulse from element $i$ and recording the scattered wave field at element $j$ (see Fig.~\ref{fig1}), for all $i$ and $j$. To perform a time-frequency analysis, these signals are filtered by a Gaussian envelope of standard deviation of 0.015 MHz centered around a given central frequency $f$. A set of matrices $\mathbf{K}(t,f)$ is obtained at each time-frequency pair. This operation was repeated for 302 different realizations of disorder by moving the array in a plane parallel to the sample surface.
\begin{figure}[htbp]
\center
\includegraphics[width=7.5cm]{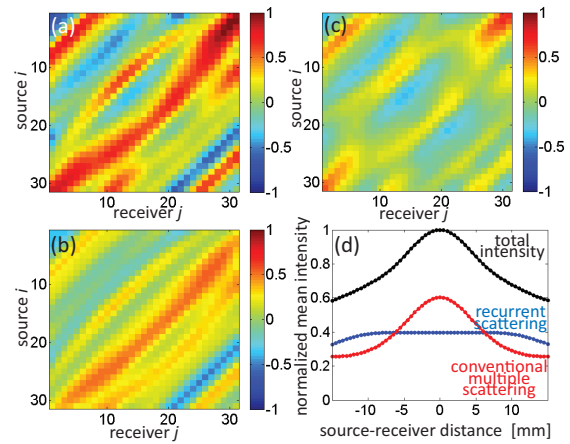}
\caption{(a) Real part of the matrix $\mathbf{K}$ at time $t=185$ $\mu$s and frequency $f=1.25$ MHz for a given realization of disorder. (b) Real part of the RS contribution $\mathbf{K_R}$. (c) Real part of the conventional MS contribution $\mathbf{K_M}$. (d) Spatial mean intensity profiles at the same time and frequency.}
\label{fig2}
\end{figure}

Figure~\ref{fig2}(a) displays a typical example of the matrix $\mathbf{K}$ obtained at frequency $f=1.25$ MHz and time $t=185$ $\mu$s much exceeding the mean free time $l_s/v_p \sim 1$ $\mu$s, deep in the MS regime. Surprisingly, despite its overall random appearance, the matrix $\mathbf{K}$ exhibits a clear coherence along its antidiagonals. This result should be considered in the context of previous studies \cite{aubry2009random,aubry_wrmc} that dealt with much weaker scattering ($kl_s \sim 100$). In these media, the antidiagonal coherence was proven to be associated with the single scattering contribution and insensitive to disorder. It can also be understood as a manifestation of the well-known memory effect.
This phenomenon was discovered in optics in the late eighties \cite{freund,feng} and has recently received renewed interest in the context of ultrasonic and optical imaging \cite{aubry_jap,putten,popoff2011,katz_2012,bertolotti}. When an incident plane wave is rotated by an angle $\theta$, the speckle image measured in transmission or in backscattering in the far-field of the sample is shifted by the same angle $\theta$ (or $-\theta$ in reflection), as long as $\theta$ does not exceed a certain threshold, namely the angular correlation width $\Delta \theta$. In the single scattering regime, the memory effect actually spreads over the whole angular spectrum ($\Delta \theta = \pi/2$) \cite{freund,feng}. This accounts for the fact that the signals $K_{ij}$ are coherent along the same antidiagonal, as long as only single scattering takes place. Indeed two pairs of array elements $(i_1,j_1)$ and $(i_2,j_2)$ on the same antidiagonal obey $i_1+j_1=i_2+j_2$. Changing the direction of emission amounts to changing $i_1$ into $i_2$. Consequently, in reflection the resulting speckle will be tilted so that the signal that was received in $j_1$ will be coherent with the new signal in $j_2=j_1-(i_2-i_1)$. When MS takes place, the correlation width $\Delta \theta$
is greatly restricted.
Therefore the $K_{ij}$'s are no longer expected to exhibit this remarkable anti-diagonal coherence, and are expected to emerge as random and only short-range correlated
\cite{aubry2009random,aubry_wrmc}. However, Fig.~\ref{fig2}(a) clearly contradicts this simple picture. It indicates that the matrix $\mathbf{K}$ corresponding to the strongly MS regime shares the spatial coherence of the matrix corresponding to the single scattering regime, though at much longer times of flight.

Another surprising result is shown by Fig.~\ref{fig2}(d) which displays the mean backscattered intensity as a function of the distance between the source $i$ and the receiver $j$ at the same time and frequency as in Fig.~\ref{fig2}(a). In the MS regime ($v_p t \gg l_s$) and far from localization ($k l_s \gg 1$), the intensity backscattered at the source is expected to be twice as large as the intensity far from the source: this is the coherent backscattering phenomenon \cite{Ishimaru1,albada,wolf,tourin}. Although we do indeed obtain this coherent backscattering peak, the enhancement factor is clearly smaller than 2. We interpret this as a sign of a large contribution from RS
(red arrows in Fig.~\ref{fig1}). 
RS, just like single scattering, contributes to the background intensity that is independent of the distance between source and receiver. The interference between a wave and its reciprocal counterpart is indeed always constructive for these two contributions. A reduction of the enhancement factor due to RS was previously observed experimentally, but in a very much lower proportion and not as a function of time \cite{wiersma2,bart}.

The very large RS contribution seen in our experiment sheds new light on the long-range spatial coherence observed in Fig.~\ref{fig2}(a). In the strongly scattering regime, the backscattered field can be decomposed into a sum of two terms:

\noindent $\bullet$ A RS contribution (red arrows in Fig.~\ref{fig1}) which displays the same statistical properties as the single scattering one. This contribution accounts for the deterministic coherence along the antidiagonals of $\mathbf{K}$ in Fig.~\ref{fig2}(a).

\noindent $\bullet$ A \textit{conventional} MS contribution (orange arrows in Fig.~\ref{fig1}) for which the first and last scattering events are separated by more than one mean free path. In this case, the memory effect is restricted to the angular width of the backscattering cone \cite{freund2,akkermans2004physique}.

Previous studies have taken advantage of the memory effect to separate single and multiple scattering \cite{aubry2009random,aubry_jap,aubry_jasa}. Here, the previous method is significantly extended \cite{supp} to enable $\mathbf{K}$ [Fig.~\ref{fig2}(a)] to be separated 
into a RS component $\mathbf{K_R}$ [Fig.~\ref{fig2}(b)] and a conventional MS component $\mathbf{K_M}$ [Fig.~\ref{fig2}(c)]. Once the separation of these two contributions is performed, one can compute the corresponding mean backscattered intensity [Fig.~\ref{fig2}(d)]. Whereas RS leads to a flat intensity profile, the conventional MS intensity exhibits a coherent backscattering cone. The recovery of an enhancement factor close to two demonstrates that RS was indeed responsible for reducing the enhancement factor seen for the total intensity $I$
\cite{bart} [Fig.~\ref{fig2}(d)].

\begin{figure}[htbp]
\center
\includegraphics[width=8.5cm]{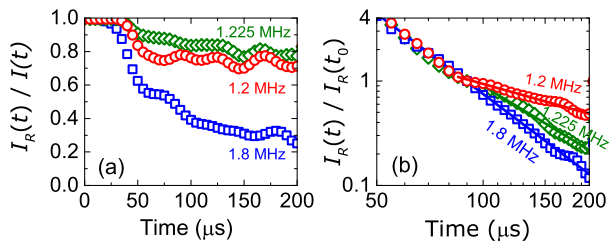}
\caption{(a) Time evolution of the RS ratio $I_R/I$ at $f=1.2$ MHz (red circles), $f=1.225$ MHz (green diamonds) and $f=1.8$ MHz (blue squares). (b) Time evolution of the RS intensities $I_R(t)$, normalized by their values at time $t_0=90$ $\mu$s,  at the same three frequencies, shown in a log-log scale. The corresponding fits with a power law (continuous lines) yield $I_R(t) \propto t^{-2.55 \pm 0.13}$ at 1.8 MHz, $I_R(t) \propto t^{-2.04 \pm 0.18}$ at 1.225 MHz and  $I_R(t) \propto t^{-0.95 \pm 0.23}$ at 1.2 MHz.}
\label{fig3}
\end{figure}
The RS intensity $I_R$ is directly related to the probability $P_R$ for a wave to return to the spot at which it entered the scattering sample. From a theoretical point of view, this return probability is a key quantity in the description of the renormalization of the diffusion constant in the SC theory of localization \cite{Vollhardt,skipetrov}. Fig.~\ref{fig3}(a) shows the time-evolution of the RS ratio $I_R/I$ at $f=1.2$ MHz (critical regime), $f=1.225$ MHz (localized regime) and $f=1.8$ MHz (diffuse regime), with $I$ the total backscattered intensity. After a plateau close to 100\% that lasts until $t=50$ $\mu$s due to specular echoes from the sample surface (see green arrows in Fig.~\ref{fig1}), the RS ratio starts to decrease with time. This ratio is highest at 1.225 MHz, and has a particularly slow decay at $f=1.2$ MHz, with $I_R/I$ being still above 70\% at $t=200$ $\mu$s (\textit{i.e} after 450 scattering events) for both frequencies. The large predominance of RS paths at frequencies near 1.225 MHz strongly suggests it to be a crucial element in any discussion of AL.

The decay of the RS intensity with time bears particular signatures of AL. For times of flight smaller than twice the Thouless time $\tau_D=L^2/D_B\sim 100$ $\mu$s, the medium can be considered as semi-infinite in a backscattering configuration. In that case, a power law decay is expected for the return probability at the surface of the sample: $I_R(t)$ should decrease as $1/t^{5/2}$ in the diffuse regime \cite{dogariu} and, for $t > \xi^2/D_B$, as $1/t^{2}$ in the localized regime \cite{skipetrov,dogariu}. At longer times of flight ($t>2\tau_D$), the finite sample size should come into play so that the decay becomes similar to that of time-dependent transmission, \textit{i.e.} exponential in the diffuse regime \cite{page2,johnson} and non-exponential in the localized regime \cite{hu,storzer}.

Figure~\ref{fig3}(b) displays the typical time dependence of $I_R(t)$ in the 1--2 MHz frequency range. At all frequencies, $I_R(t)$ can be described by a power law $1/t^{\alpha}$ between 85 and 200 $\mu$s. We recover the exponent $\alpha = 5/2$ at 1.8 MHz, characteristic of the diffuse regime, and observe its shift to $2$ at 1.225 MHz as expected in the localized regime. Furthermore, we observe a strikingly slower decay at $f=1.2$ MHz, which corresponds to a mobility edge and where $\alpha$ reaches a value close to $1$. Further theoretical work on the Anderson transition in open media is needed to account for this unexpected behavior. As the exponent $\alpha$ is linked to the dimension of available space in which the waves propagate, it might be related to previously observed \cite{faez} and theoretically predicted \cite{wegner,evers08,chalker,kravtsov} multifractal properties of the wave function. In the literature, transmission measurements have also shown some deviations from SC theory in the localized regime \cite{zhang} due to long-lived resonant modes \cite{tian}. Interestingly, as discussed below, the discrepancy with SC theory in backscattering coincides with the emergence of very intense RS paths close to the mobility edge.

\begin{figure}[htbp]\center
\includegraphics[width=8.5cm]{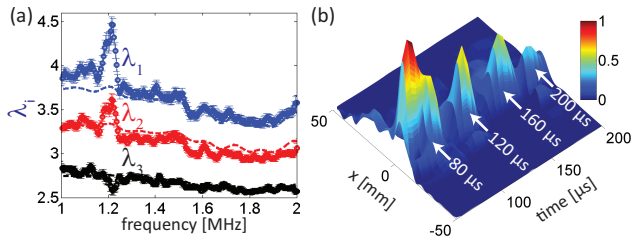}
\caption{(a) Average of the first three singular values $\left \langle \lambda_i \right \rangle$ versus frequency at time $t=150$ $\mu$s. The experimental results (circles) are compared to the RMT predictions (lines). Error bars correspond to $\pm$ the standard error in the mean. (b) Backpropagation of the first singular vector at the sample surface computed at each time for one realization of disorder at frequency $f=1.2$ MHz. $x$ represents the coordinate along the surface.}
\label{fig4}
\end{figure}

We now investigate the effect of RS on the statistical properties of the random matrix $\mathbf{K}$. First, we study the statistical properties of the singular values of $\mathbf{K}$ and compare them with the predictions of random matrix theory (RMT). The singular value decomposition (SVD) consists in writing $\mathbf{K}=\mathbf{U}\mathbf{\Lambda}\mathbf{V}^{\dag}$, where $\mathbf{U}$ and $\mathbf{V}$ are unitary matrices and $\mathbf{\Lambda}$ is a diagonal matrix whose non-zero elements $\lambda_i$ are called the singular values of $\mathbf{K}$. They are real, positive and arranged in decreasing order. In the case of random matrices whose antidiagonals are either constant or strongly correlated (Hankel matrices), the statistical distribution of singular values, particularly the strongest ones, can be calculated and compared to experimental measurements \cite{aubry2009random,aubry_wrmc}. In the present situation, given the substantial memory effect, one would expect the singular values to follow Hankel-like distributions, as was observed in previous work dealing with weaker scattering \cite{aubry2009random,aubry_wrmc}. However, the present experimental results show that this is clearly not the case for the first two singular values, especially around 1.2 MHz. Figure~\ref{fig4}(a) displays the average of the three largest singular values, $\left \langle \lambda_{i=1,2,3} \right \rangle$, as a function of frequency at a given time $t=150$ $\mu$s \cite{footnote1}. The $\left \langle \lambda_i \right \rangle$ are compared to the theoretical values expected for Hankel random matrices
\cite{aubry_wrmc}. 
As we will now discuss, the spectacular discrepancy between the first two singular values and RMT predictions around 1.2 MHz is related to the dominance of intense RS paths in the backscattered wave field.

Singular vectors can be given a physical interpretation: they are the invariants of the time reversal operator $\mathbf{K}\mathbf{K^{\dag}}$.
In a weakly scattering regime, there is a one-to-one correspondence between single scattering paths and eigenvectors of $\mathbf{K}$ associated with non-zero singular values \cite{prada1994time,popoff2}: each singular vector of $\mathbf{K}$ corresponds to the wave front that, if sent as such from the array, would focus onto the corresponding scatterer. By contrast, in strongly scattering media, the physical meaning of singular vectors and the result of their backpropagation will be different, and may reveal unusual phenomena.
Figure~\ref{fig4}(b) shows the time evolution of the numerical backpropagation of $\mathbf{V_1}$ at the sample surface for one realization of disorder. Once again, we observe remarkable behavior around 1.2 MHz: $\mathbf{V_1}$ back-focuses on the same particular location at regular time intervals (every 40 $\mu$s), as if it were associated with a particular \textit{dynamic} \textit{hotspot}. The corresponding peaks are of same duration as the incident pulse ($\sim$ 10 $\mu$s). In the strongly scattering regime, near the localization transition, we argue that the largest singular values of $\mathbf{K}$ are associated with predominant RS paths, whose entry/exit points appear as specific hotspots at the surface of the sample. For the realization of disorder considered in Fig.~\ref{fig4}(b), the occurrence of the same hot spot every 40 $\mu$s 
could indicate that it corresponds to successive round trips along a RS path of no less than 90 scattering events. Note that the emergence of dynamic hot spots has been observed in other configurations of disorder with different regularly-spaced intervals of time ranging from 30 $\mu$s to 100 $\mu$s \cite{supp}. From a practical point of view, the selective and independent excitation of RS paths can open new perspectives for manipulation of wave fields in complex media. In an amplifying medium, for example, one could select the scattering paths to be amplified and thus control the random laser process \cite{bachelard}.


In conclusion, we have shown how new information on the dynamics of Anderson localization can be obtained using the mesoscopic reflection matrix approach. The long-range spatial coherence of the RS contribution has been directly observed and its link with the memory effect has been established.   A sophisticated matrix manipulation method has been further developed to separate this contribution from the conventional MS background, enabling RS to be examined experimentally on its own for the first time. Thus we are able not only to demonstrate that the contribution of RS to the total backscattered intensity is strikingly large near the Anderson transition, but also to investigate the dynamics of the return probability for waves within the scattering medium.  In particular, a dramatic slowing down in the decay of the return probability has been found near the mobility edge, with a power-law exponent $\alpha$ that is smaller than expected, motivating new theoretical work to understand this behavior quantitatively.  The emergence of very intense RS paths, which are revealed by our analysis of the singular values of the backscattering matrix, is another novel feature of our results, and provides support for the idea that RS plays a very important role near the mobility edge.

A.A. and A.D. are grateful for funding provided by the Agence Nationale de la Recherche (ANR-11-BS09-007-01, Research Project DiAMAN) and by LABEX WIFI (Laboratory of Excellence ANR-10-LABX-24) within the French Program Investments for the Future under reference ANR-10- IDEX-0001-02 PSL?. J.H.P. and L.A.C. acknowledge the support of NSERC  (Discovery Grant RGPIN/9037-2011, Canada Government Scholarship, and Michael Smith Foreign Study Supplement), the Canada Foundation for Innovation and the Manitoba Research and Innovation Fund (CFI/MRIF, LOF Project 23523).  

%
%

\begin{thebibliography}{44}
\bibitem[]{footnote1}
\bibinfo{note}{Prior to averaging over different realizations of disorder, the singular values have been normalized by their quadratic mean.}
\end{thebibliography}
\end{document}